\documentclass[preprints,article,accept,pdftex,moreauthors]{mdpi} 
\firstpage{1} 
\pubvolume{1}
\issuenum{1}
\articlenumber{0}
\pubyear{2025}
\copyrightyear{2025}
\datereceived{ } 
\daterevised{ } % Comment out if no revised date
\dateaccepted{ } 
\datepublished{ } 
\hreflink{https://doi.org/} % If needed use \linebreak
	\usepackage{GB_packages}
	\usepackage{GB_glossaries}
	
	\Title{Influence of Uncertainties in Optode Positions on Self-Calibrating or Dual-Slope Diffuse Optical Measurements}
	\TitleCitation{Influence of Uncertainties in Optode Positions on Self-Calibrating or Dual-Slope Diffuse Optical Measurements}
	
	\Author{Giles Blaney *\orcidA{}, Angelo Sassaroli \orcidB{}, Tapan Das \orcidC{}, and Sergio Fantini \orcidD{}}
	\AuthorNames{Giles Blaney, Angelo Sassaroli, Tapan Das, and Sergio Fantini}
	\AuthorCitation{Blaney, G.; Sassaroli, A.; Das, T.; Fantini, S.}
	\address[1]{Tufts University, Department of Biomedical Engineering, 4 Colby Street, Medford MA, USA, 02155}
	\corres{Correspondence: Giles.Blaney@tufts.edu}
	
	\abstract{%200/200
		Self-calibrating and dual-slope measurements have been used in the field of diffuse optics for robust assessment of absolute values or temporal changes in the optical properties of highly scattering media and biological tissue.
		These measurements employ optical probes with a minimum of two source positions and a minimum of two detector positions. 
		This work focuses on a quantitative analysis of the impact of errors in these source and detector positions on the assessment of optical properties. 
		We considered linear, trapezoidal, and rectangular optode arrangements and theoretical computations based on diffusion theory for semi-infinite homogeneous media. 
		We found that uncertainties in optodes' positions may have a greater impact on measurements of absolute scattering versus absorption coefficients. 
		For example, \SI{4.1}{\percent} and \SI{19}{\percent} average errors in absolute absorption and scattering, respectively, can be expected by displacing every optode in a linear arrangement by \SI{1}{\milli\meter} in any direction.
		The impact of optode position errors is typically smaller for measurements of absorption changes. 
		For each geometrical arrangement (linear, trapezoid, rectangular), we identify the direction of the position uncertainty for each optode that has minimal impact on the optical measurements. 
		These results can guide the optimal design of optical probes for self-calibrating and dual-slope measurements.
	}
	
	
\begin{document}
		%%%%%%%%%%%%%%%%%%%%%%%%%%%%%%%%%%%%%%%%%%
		\section{Introduction}
		In \gls{FD} \gls{NIRS}, strategies for the retrieval of the absolute optical properties are based on multi-frequency \cite{Haskell_JOSAA94_BoundaryConditions} or multi-distance \cite{Fantini_JOSAB94_SemiinfinitegeometryBoundary} measurements. 
		In both cases, a calibration of the experimental apparatus on a phantom of known optical properties is required. 
		This is not a trivial step, and the success of the calibration procedure for the quantification of the absolute optical properties of a target medium is based on several assumptions: 
		1) that a calibration phantom of known (i.e., high accuracy) optical properties and robust procedures for its quantification exist \cite{Pifferi_AO05_PerformanceAssessment}; 
		2) that the phantom is optically homogeneous; 
		3) that one is able to reproduce the optical coupling between the probe and the calibration phantom also during the experiment on a medium of unknown optical properties \cite{Fantini_OE95_FrequencydomainMultichannel}; 
		4) that the laser power and the detector gain are stable. 
		These are not conditions that are always easy to satisfy. 
		For example, experiments that last several hours might need recalibration of the system due to laser power and detector gain fluctuations. 
		Ideally, pressure sensors to control the optical coupling might be needed. 
		During in vivo experiments on human subjects, the requirement of stable coupling might be particularly difficult to achieve, due to uncontrolled subject’s motion. 
		Therefore, methods that bypass the limiting factors of calibration and that provide accurate estimates of the optical properties are highly welcome.
		\par
		
		Hueber et al.\ \cite{Hueber_Opt.Tomogr.Spectrosc.TissueIII99_NewOptical} first proposed the \gls{SC} method for the retrieval of the absolute optical properties of tissues in \gls{FD} \gls{NIRS} (namely the \gls{mua}, and the \gls{musp}). 
		The proposed method provided a robust way to retrieve the optical properties with the advantage of being highly insensitive to the limiting factors intrinsic in a calibration procedure. 
		Since then, the \gls{SC} method has found a number of applications for the estimation of \gls{StO2} \cite{Franceschini_OL99_NoninvasiveOptical, MacLeod_JCVA12_DevelopmentValidation, Kleiser_JBO18_VivoPrecision, Wu_NI22_OpensourceFlexNIRS}, for the measurement of \gls{dmua} of target tissues during protocols that elicit hemodynamic changes \cite{Xu_AO07_DevelopmentHandheld}, and spectral measurements of the absolute optical properties of tissues \cite{Blaney_AO21_BroadbandAbsorption,Pham_BOE24_CrosswavelengthCalibrating}.
		When implemented to recover \gls{dmua} from a single type of optical data (e.g., only the \gls{phi}\glsreset{phi} in \gls{FD}), the method is referred to as \gls{DS}, which can been proposed in \gls{FD} \cite{Sassaroli_JOSAA19_DualslopeMethod} or \gls{TD} \cite{Sawosz_BOE19_MethodImprove} and implemented in \gls{FD} \cite{Blaney_JBP20_PhaseDualslopes}.
		The basic ``unit'' for \gls{SC} measurement features two sources and two detectors in a special geometrical arrangement. 
		The four optodes identify four source-detector separations, and the prerequisite of a \gls{SC} unit is that the two short and the two long source-detector separations must be equal.
		If this requirement is met (with some spatial constraints due to the dynamic range of detectors), one is free to choose between several spatial arrangements of the optodes.
		Typical ones are the linear (i.e. with the detectors sandwiched between the sources along a line), rectangular and trapezoidal (i.e., the optodes occupy the vertices of a rectangle and a trapezoid, respectively) \cite{Fantini_JIOHS19_TransformationalChange}.
		Examples of these various arrangements can be found in literature: 
		linear \cite{Hueber_Opt.Tomogr.Spectrosc.TissueIII99_NewOptical,Franceschini_OL99_NoninvasiveOptical,Davie_Anesthesiology12_ImpactExtracranial,MacLeod_JCVA12_DevelopmentValidation,Scholkmann_PM14_MeasuringTissue,Kleiser_JBO18_VivoPrecision,Fantini_JIOHS19_TransformationalChange,Sassaroli_JOSAA19_DualslopeMethod,Sawosz_BOE19_MethodImprove,Blaney_Rev.SI20_DesignSource,Blaney_JBP20_PhaseDualslopes,Blaney_AO21_BroadbandAbsorption,Blaney_MDPIAppSci21_DualslopeDiffuse,Perekatova_MDPIDia.23_VISNIRDiffuse,Pham_BOE24_CrosswavelengthCalibrating};
		rectangular \cite{Hueber_Opt.Tomogr.Spectrosc.TissueIII99_NewOptical,Xu_AO07_DevelopmentHandheld,Jenny_JBO11_ReproducibilityCerebral,Fantini_JIOHS19_TransformationalChange,Wu_NI22_OpensourceFlexNIRS,Wu_NPh23_EnhancingDiffuse,Frias_NeuralImagingSens.202424_FunctionalBrain}; and 
		trapezoidal \cite{Chincarini_MDPIAni.19_EvaluationSheep,Fantini_JIOHS19_TransformationalChange,Blaney_Rev.SI20_DesignSource,Frias_NeuralImagingSens.202424_FunctionalBrain}.
		\par
		
		Since its inception, the \gls{SC} method has been also used with \gls{CW} instrumentation.
		For example, Jenny et al.\ \cite{Jenny_JBO11_ReproducibilityCerebral} developed a custom-made \gls{CW} system with a probe having two concentric rectangular \gls{SC} units and three wavelengths at each source location.
		The instrument was used for assessing the precision of \gls{StO2} measurements in vivo on a neonatal population. 
		The same instrument was modified by Kleiser et al.\ \cite{Kleiser_BOE16_ComparisonTissue} in a custom-made \gls{CW} instrument (OxyPrem v. 1.3) featuring two collinear \gls{SC} units, i.e., having two detectors and four source locations (four wavelengths per source location) sandwiched between the detectors in a linear array.
		The instrument was used to determine the accuracy of \gls{StO2} measurements on a blood-lipid phantom.
		Kleiser et al.\ \cite{Kleiser_JBO18_VivoPrecision} used the same instrument for \gls{StO2} precision assessment in a neonatal population. 
		Chincarini et al.\ \cite{Chincarini_MDPIAni.19_EvaluationSheep} used OxyPrem v. 1.4 featuring two \gls{SC} units in a trapezoidal arrangement, and measured changes of oxy- and deoxyhemoglobin on sheep.
		Raoult et al.\ \cite{Raoult_MDPIAni.18_ValenceIntensity} used OxyPrem for measuring changes in oxy- and deoxyhemoglobin in dogs.
		Another commercial instrument, the Nonin EQUANOX, has used a probe with the \gls{SC} arrangement to measure cerebral oxygenation in humans \cite{Davie_Anesthesiology12_ImpactExtracranial}, but it is unclear if the \gls{SC} method was implemented in the data analysis.
		Xu et al.\ \cite{Xu_AO07_DevelopmentHandheld} developed a custom-made \gls{CW} system with \num{8} linear \gls{SC} arrangement (two wavelength per source location) combined in a hand-held probe for dynamic characterization of biological tissues.
		Scholkmann et al.\ \cite{Scholkmann_PM14_MeasuringTissue}, used OxyPrem to test the resilience of measurements towards induced motion artifacts.
		Perekatova et al.\ \cite{Perekatova_MDPIDia.23_VISNIRDiffuse} used a custom-built \gls{CW} ultra-broadband (\SIrange{460}{1030}{\nano\meter}) featuring a \gls{SC} linear unit and retrieved the \gls{mueff} at different wavelengths.
		The experimental setup was tested for resistance toward different instrumental perturbations.
		Wu et al.\ \cite{Wu_NI22_OpensourceFlexNIRS} developed a \gls{CW} system with a probe comprising a \gls{SC} unit in a rectangular arrangement (with two wavelengths per source location).
		The system was used for studying pulsatile cerebral blood flow together with a \gls{DCS} system \cite{Wu_NPh23_EnhancingDiffuse}.
		\par
		
		We note that when \gls{CW} instrumentation is used for measuring \gls{StO2} or \gls{dmua}, one must make assumptions about the values and wavelength dependence of the reduced scattering coefficient.
		This source of error and how it propagates to the target parameters has not been investigated. 
		Also, there has not been a thorough investigation into the errors on the source-detector separations and how it propagates to the target parameters. 
		This error was briefly addressed by Hueber et al.\ \cite{Hueber_Opt.Tomogr.Spectrosc.TissueIII99_NewOptical} for the linear arrangement by considering: 
		a) errors in only one short distance; 
		b) errors in all the short distances; 
		c) errors in all short and long distances (by keeping the difference of the distances unchanged); 
		d) errors in all short and long distances by keeping the average distance unchanged. 
		In this work, we consider a wider range of cases for the errors in the source-detector separation and their propagation for the quantification of absolute optical properties (i.e., \gls{mueff} or \gls{mua} \& \gls{musp}), or absorption changes (i.e., \gls{dmua}). 
		We also consider four typical \gls{SC} arrangements: linear, asymmetric-linear, trapezoidal, diagonal-rectangular.
		\par
		
		%%%%%%%%%%%%%%%%%%%%%%%%%%%%%%%%%%%%%%%%%%
		\section{Materials and Methods}
		\subsection{Dual-slope Arrangements}
\begin{figure}[H]
			\includegraphics{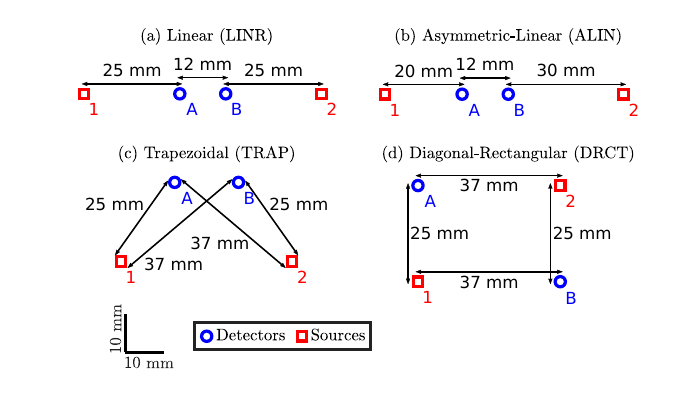}
			\caption{The four types of \gls{DS} arrangement considered in this work. Sources are shown as red squares labeled with numbers and detectors are blue circles labeled with letters. Distances are measured from optode center to optode center.
				(a) The LINeaR (LINR) arrangement with \glspl{rho} of $[25, 25, 37, 37]~\si{\milli\meter}$.
				(b) The Asymmetric-LINear (ALIN) arrangement with \glspl{rho} of $[20, 30, 32, 42]~\si{\milli\meter}$.
				(c) The TRAPezoidal (TRAP) arrangement with \glspl{rho} of $[25, 25, 37, 37]~\si{\milli\meter}$.
				(d) The Diagonal-ReCTangular (DRCT) arrangement with \glspl{rho} of $[25, 25, 37, 37]~\si{\milli\meter}$.
				\label{fig:geo}}
		\end{figure}

		In this work, four types of \gls{DS} arrangement geometry will be considered (\autoref{fig:geo}).
		The first arrangement, LINeaR (LINR; \autoref{fig:geo}(a)), is the simplest and most common type of \gls{DS} set.
		The next arrangement, Asymmetric-LINear (ALIN; \autoref{fig:geo}(b)), is a slight modification on LINR with the detectors shifted to the left.
		ALIN has not been implemented in as many experiments as the other sets, but has been shown to have a similar region of spatial sensitivity as LINR \cite{Fantini_JIOHS19_TransformationalChange}.
		The last two sets, TRAPezoidal (TRAP; \autoref{fig:geo}(c)) and Diagonal-ReCTangular (DRCT; \autoref{fig:geo}(c)), are the two types of sets found in the modular hexagonal imaging arrangement proposed by our group \cite{Frias_NeuralImagingSens.202424_FunctionalBrain}.
		TRAP particularly is of interest since most \gls{DS} imaging arrays are comprised of mainly trapezoidal sets \cite{Blaney_Rev.SI20_DesignSource}.
		All of these arrangements have the same mean \gls{rho} of \SI{31}{\milli\meter}.
		\par
		
		\glsreset{rho}
		
		\subsection{Analytical Forward Model}
		We generated forward \gls{FD} \gls{NIRS} data in the form of the \gls{RCom} for each source-detector pair in the arrangement being assessed (i.e., 1A, 1B, 2A, and 2B).
		These \gls{RCom} data were computed using an analytical diffusion model for a semi-infinite homogeneous medium \cite{Bigio_16_QuantitativeBiomedical, Haskell_JOSAA94_BoundaryConditions}.
		The primary inputs to the forward model are \gls{mua}, \gls{musp}, and \gls{rho}, of which the former two were varied and the latter one was determined by the arrangement.
		The remaining model parameters were the \gls{fmod} and the \gls{n} which were fixed at \SI{100e6}{\hertz} and \num{1.4}, respectively.
		Importantly, when considering an optode position error, data were generated with \glspl{rho} corresponding to the optode in the error position not the nominal position.
		\par
		
		We also considered changes in \gls{RCom} resulting from small changes in \gls{mua} to test \gls{DS}'s ability to recover \glspl{dmua}.
		These were computed by finding \gls{RCom} for the baseline optical properties and for the same parameters except for \gls{mua} that was increased by \SI{0.0001}{\per\milli\meter} (i.e., true \gls{dmua} of \SI{0.0001}{\per\milli\meter}).
		\par
		
		\subsection{Inverse Models}
		\subsubsection{Absolute Optical Properties}
		For a given simulation case, the forward model generated four values of \gls{RCom}, one for each source-detector pair in the arrangement of interest.
		These four \glspl{RCom} are input into the inverse models for recovering the \gls{mua} and the \gls{musp} to implement the \gls{SC} method.
		The remaining parameters which must be assumed for the inverse models were the \glspl{rho}, the \gls{fmod}, and the \gls{n}.
		The \glspl{rho} were either assigned the correct values (given the optode position considered in the forward model) or incorrect values.
		The cases where the \glspl{rho} were assumed incorrectly correspond to cases where optodes were not in their nominal position to generate data with the forward model, but the nominal \glspl{rho} were assumed in the inverse model.
		This represents a possible real-life scenario where the optode arrangement was affected by position errors, but the data were analyzed assuming the nominal \glspl{rho}.
		The remaining two parameters, the \gls{fmod} and the \gls{n}, were assumed to be known in all cases.
		\par
		
		In this work, we have tested two inverse models based on the \gls{SC} \gls{RCom} data to recover the \gls{mua} and the \gls{musp}.
		We will refer to the first method as the ``slopes'' method since it is based on the slopes, computed with \gls{SC} / \gls{DS} methods, of the \gls{lnr2I} and the \gls{phi} with respect to \gls{rho} \cite{Fantini_PMB99_NoninvasiveOptical}.
		The second method we will refer to as the ``iterative'' method since it is based on iteratively solving the forward model for \gls{RCom} as a function of \gls{rho} \cite{Blaney_MDPIAppSci21_DualslopeDiffuse}, where $\as{RCom}=\as{I}e^{i\as{phi}}$.
		The ``slopes'' method solves the following equation:
		\begin{equation}
			\ln\left[ \as{rho}^2 \as{RCom} \right] = -\as{mueffCom} \as{rho}
		\end{equation}
		where the \gls{mueffCom} is defined as:
		\begin{equation}
			\as{mueffCom}=\sqrt{3(\as{musp}+\as{mua})(\as{mua}-2\pi\as{fmod} i /v)}
		\end{equation}
		where $v$ is the speed of light in the medium.
		The ``iterative'' method, on the other-hand, iteratively solves this expression:
		\begin{equation}
			\ln\left[ \widetilde{f}(\as{rho},\as{mua},\as{musp}) \as{RCom} \right] = -\as{mueffCom} \sqrt{\as{rho}^2+1/\as{musp}^2}
		\end{equation}
		by updating the \gls{mua} and the \gls{musp} at every iteration. 
		Details on the iterative method and the complex function $\widetilde{f}(\as{rho},\as{mua},\as{musp})$ can be found in Reference~\citenum{Blaney_MDPIAppSci21_DualslopeDiffuse}'s Equation~(4).
		The slope method is based on stricter assumptions than the iterative method but is less computationally intense and more stable against artifacts in noisy data.
		\par
		
		\subsubsection{Changes in Absorption}
		The forward model considered changes in the \gls{mua} by generating \glspl{RCom} for both the baseline optical properties and perturbed optical properties for each of the four source-detector pairs in the \gls{DS} arrangement of interest.
		From these baseline and perturbed \glspl{RCom}, the changes in the \gls{DS} of \gls{lnI} (i.e., $\ln(|\as{RCom}|)$) or \gls{phi} (i.e., $\angle \as{RCom}$) were calculated and converted to \gls{dmua} using previously reported methods \cite{Blaney_JBP20_PhaseDualslopes, Sassaroli_JOSAA19_DualslopeMethod}.
		It is worth noting that these methods require an assumption of \gls{mua} and \gls{musp} to calculate the \glspl{LpathCom}.
		The values of \gls{mua} and \gls{musp} found using the iterative method on the forward data were used in this case, meaning errors in the recovered \gls{mua} and \gls{musp} may propagate through \glspl{LpathCom} to an error on the recovered \gls{dmua}.
		The inverse model for absorption changes also requires an assumption of the \glspl{rho}.
		These \glspl{rho} were assumed in the same way as described for the absolute optical property inverse model, with the nominal \glspl{rho} always used in the inverse problem even when an error is imposed on the optode positions to generate the data with the forward model.
		\par
		
		\subsection{Variable Definitions}
		The results in this work focus on errors in recovered parameters as a result of incorrect assumed values of \gls{rho} due to geometrical optode position errors.
		Therefore, we must be careful to use a notation which differentiates true parameters from parameters recovered either with correct or incorrect assumptions.
		Accordingly, variables with $(\surd\as{rho})$ in superscript indicate values recovered by the inverse model assuming the correct \glspl{rho} while variables with $(\times\as{rho})$ in superscript indicate values recovered assuming the incorrect \glspl{rho}.
		Variables without either $(\surd\as{rho})$ or $(\times\as{rho})$ in superscript indicate the true values used in the forward model to generate simulated data.
		For example, \as{mua} indicates the true \acrlong{mua}\glsunset{mua} used in the forward model while $\mu_a^{(\surd\as{rho})}$ or $\mu_a^{(\times\as{rho})}$ indicate the \acrlong{mua} recovered by the inverse model assuming the correct or incorrect \glspl{rho}, respectively.
		\par
		
		We also aim to assess to what extent an error in an optode's position effects the values recovered in general.
		To quantify this we define the root-mean-square-error in either the \gls{mua} or the \gls{musp} from varying the optode positions in a \SI{2}{\milli\meter} diameter circle which orbits their nominal position ($\sigma^{\varnothing\SI{2}{\milli\meter}}_{\as{mua}}$ or $\sigma^{\varnothing\SI{2}{\milli\meter}}_{\as{musp}}$, respectively) as follows:
		\begin{equation}\label{equ:RMSEmua}
			\sigma^{\varnothing\SI{2}{\milli\meter}}_{\as{mua}}=\sqrt{\frac
				{\sum_{i=1}^{n^{\varnothing\SI{2}{\milli\meter}}} \left( \mu_{a,i}^{(\times\rho)} - \mu_a^{(\surd\rho)} \right)^2 }
				{n^{\varnothing\SI{2}{\milli\meter}}}}
		\end{equation}
		\begin{equation}\label{equ:RMSEmusp}
			\sigma^{\varnothing\SI{2}{\milli\meter}}_{\as{musp}}=\sqrt{\frac
				{\sum_{i=1}^{n^{\varnothing\SI{2}{\milli\meter}}} \left( \mu_{s,i}^{\prime,(\times\rho)} - \mu_s^{\prime,(\surd\rho)} \right)^2 }
				{n^{\varnothing\SI{2}{\milli\meter}}}}
		\end{equation}
		Where, $\mu_a^{(\surd\rho)}$ and $\mu_s^{\prime,(\surd\rho)}$ are the \acrlong{mua}\glsunset{mua} and the \acrlong{musp}\glsunset{musp} recovered by the inverse model with the optode in its nominal position and assuming the nominal \glspl{rho}.
		However, $\mu_a^{(\times\rho)}$ and $\mu_s^{\prime,(\times\rho)}$ are the \acrlong{mua} and the \acrlong{musp} recovered by the inverse model with the optode \emph{not in its nominal position but assuming the nominal} \glspl{rho}.
		To realize this, the optode's (e.g., source 1's) position is varied through $n^{\varnothing\SI{2}{\milli\meter}}$ points in a \SI{2}{\milli\meter} diameter circle surrounding its nominal position (e.g., a circle orbiting source 1's nominal position with a \SI{1}{\milli\meter} radius) and computing the forward then inverse models for each point.
		To assess the influence of a position error in multiple optodes (e.g., source 1 and detector A) each optode considers $\sqrt[m]{n^{\varnothing\SI{2}{\milli\meter}}}$ positions (e.g., $\sqrt{n^{\varnothing\SI{2}{\milli\meter}}}$) in a circle around its nominal position, where $m$ is the number of optodes considered (e.g., \num{2}).
		These positions are co-varied for each optode resulting in $n^{\varnothing\SI{2}{\milli\meter}}$ values of $\mu_a^{(\times\rho)}$ and $\mu_s^{\prime,(\times\rho)}$ which may be used to compute $\sigma^{\varnothing\SI{2}{\milli\meter}}_{\as{mua}}$ or $\sigma^{\varnothing\SI{2}{\milli\meter}}_{\as{musp}}$ with \autoref{equ:RMSEmua} and \autoref{equ:RMSEmusp}.
		It should be noted that this definition quantifies deviation from the recovered nominal value instead of deviation from the true value to remove any effect of systematic error in the inversion.

		%%%%%%%%%%%%%%%%%%%%%%%%%%%%%%%%%%%%%%%%%%
		\section{Results}
		\subsection{Self-Calibrated Recovery of Absolute Optical Properties}\label{sec:RESsc}
		\subsubsection{Error from the Choice of Optical Property Recovery Method}
		We have compared the accuracy of two methods to recover the \gls{mua} and the \gls{musp} from \gls{SC} \gls{FD} \gls{NIRS} data.
		These two methods are the slopes method \cite{Fantini_PMB99_NoninvasiveOptical} and the iterative method \cite{Blaney_MDPIAppSci21_DualslopeDiffuse}, where the slope method is simpler but contains stronger assumptions in its derivation.
		\autoref{tab:absMethAcc} shows the accuracy of $\mu_a^{(\surd\rho)}$ and $\mu_s^{\prime,(\surd\rho)}$ recovered using each method for a range of true values of \gls{mua} and \gls{musp}.
		We conducted this exercise for two sets of \glspl{rho}.
		The first set, $\as{rho}=[25,25,37,37]~\si{\milli\meter}$, corresponds to the LINR, TRAP, and DRCT sets (\autoref{fig:geo}(a)(c)(d)).
		The second set, $\as{rho}=[20,30,32,42]~\si{\milli\meter}$, corresponds to the ALIN set (\autoref{fig:geo}(b)).
		\par
		
		\begin{table}[H] 
			\caption{Comparison of the accuracy of the slopes \cite{Fantini_PMB99_NoninvasiveOptical} versus the iterative \cite{Blaney_MDPIAppSci21_DualslopeDiffuse} recovery method \label{tab:absMethAcc}}
			\begin{tabularx}{\linewidth}{
					c|
					S[table-format=1.3]S[table-format=1.1]|
					S[table-format=1.3]S[table-format=1.3]|
					S[table-format=1.6]S[table-format=1.4]
				}
				\toprule
				& 
				& 
				& 
				\multicolumn{4}{c}{\textbf{Recovery Method}} \\
				
				\textbf{\acrshortpl{rho}\textsuperscript{\textasteriskcentered}} & 
				\textbf{\acrshort{mua}} & 
				\textbf{\acrshort{musp}} & 
				\multicolumn{2}{c|}{\textbf{Slopes}} &
				\multicolumn{2}{c}{\textbf{Iterative}} \\
				
				\textbf{(\si{\milli\meter})} & 
				\textbf{(\si{\per\milli\meter})} & 
				\textbf{(\si{\per\milli\meter})} & 
				\textbf{$\frac{\mu_a^{(\surd\rho)}-\acrshort{mua}}{\acrshort{mua}}$ (\si{\percent})} & 
				\textbf{$\frac{\mu_s^{\prime,(\surd\rho)}-\acrshort{musp}}{\acrshort{musp}}$ (\si{\percent})} & 
				\textbf{$\frac{\mu_a^{(\surd\rho)}-\acrshort{mua}}{\acrshort{mua}}$ (\si{\percent})} & 
				\textbf{$\frac{\mu_s^{\prime,(\surd\rho)}-\acrshort{musp}}{\acrshort{musp}}$ (\si{\percent})} \\
				
				\midrule
				%\multirow{9}{*}{{[}25, 25, 37, 37{]}}
				%		\parbox[t]{2mm}{\multirow{9}{*}{\rotatebox[origin=c]{90}{{[}25, 25, 37, 37{]}}}}
				\multirow{9}{*}{\rotatebox[origin=c]{90}{{[}25, 25, 37, 37{]}}}
				& 0.005 & 0.5 & 11 & -12 & 0.024 & -0.99 \\ 
				& 0.005 & 1.0 & 9.3 & -2.9 & 0.0058 & -0.50 \\ 
				& 0.005 & 1.5 & 7.0 & -1.2 & 0.0013 & -0.33 \\ 
				& 0.010 & 0.5 & 5.0 & -12 & 0.029 & -2.0 \\ 
				& 0.010 & 1.0 & 4.7 & -3.5 & 0.0038 & -0.99 \\ 
				& 0.010 & 1.5 & 3.6 & -1.7 & 0.0046 & -0.67 \\ 
				& 0.015 & 0.5 & 2.9 & -13 & 0.037 & -3.0 \\ 
				& 0.015 & 1.0 & 3.1 & -4.0 & 0.0040 & -1.5 \\ 
				& 0.015 & 1.5 & 2.4 & -2.1 & 0.0025 & -1.00 \\ 
				\midrule
				%		\multirow{9}{*}{{[}20, 30, 32, 42{]}}
				\multirow{9}{*}{\rotatebox[origin=c]{90}{{[}20, 30, 32, 42{]}}}
				& 0.005 & 0.5 & 11 & -13 & 0.025 & -1.00 \\ 
				& 0.005 & 1.0 & 9.7 & -3.2 & 0.0070 & -0.50 \\ 
				& 0.005 & 1.5 & 7.4 & -1.3 & 0.0016 & -0.33 \\ 
				& 0.010 & 0.5 & 4.7 & -13 & 0.032 & -2.0 \\ 
				& 0.010 & 1.0 & 4.9 & -3.8 & 0.0046 & -0.99 \\ 
				& 0.010 & 1.5 & 3.8 & -1.8 & 0.0011 & -0.66 \\ 
				& 0.015 & 0.5 & 2.6 & -14 & 0.041 & -3.0 \\ 
				& 0.015 & 1.0 & 3.3 & -4.3 & 0.0048 & -1.5 \\ 
				& 0.015 & 1.5 & 2.6 & -2.2 & 0.0030 & -1.00 \\ 
				\bottomrule
			\end{tabularx}
			\\[1ex]
			\noindent{\footnotesize{Symbols: \Acrfull{rho}, \acrfull{mua}, \acrfull{musp},
					true optical property of the medium ($\mu$),
					optical property recovered with the optode in the nominal position ($(\surd\rho)$ in superscript)}}\\
			\noindent{\footnotesize{\textsuperscript{\textasteriskcentered}} {[}25, 25, 37, 37{]} corresponds to the LINR, TRAP, and DRCT arrangements; {[}20, 30, 32, 42{]} corresponds to the ALIN arrangement.}
		\end{table}
		
		Overall, the results in \autoref{tab:absMethAcc} show that the slopes method is less accurate than the iterative method.
		To summarize, averaging the error in $\mu_a^{(\surd\rho)}$ for all optical properties and \gls{rho} sets yields \SI{6}{\percent} and \SI{0.01}{\percent}, for the slopes and iterative method, respectively.
		Similarly, averaging the error in $\mu_s^{\prime,(\surd\rho)}$ for all optical properties and \gls{rho} sets yields \SI{-6}{\percent} and \SI{-1}{\percent}, for the slopes and iterative method, respectively.
		Both methods have a positive bias for $\mu_a^{(\surd\rho)}$ and a negative bias for $\mu_s^{\prime,(\surd\rho)}$.
		In either case, the accuracy of the method depends strongly on the optical properties and only depends weekly on the set of \glspl{rho}, so we will focus on interpreting the results for the $\as{rho}=[25,25,37,37]~\si{\milli\meter}$ set of \glspl{rho} here.
		The accuracy of the $\mu_a^{(\surd\rho)}$ recovered by the slopes method depends heavily on the true \gls{mua}, with worse accuracies occurring at low \glspl{mua}.
		Meanwhile, the accuracy of the $\mu_a^{(\surd\rho)}$ recovered by the iterative method more strongly depends on the true \gls{musp}, and only weakly depends on the true \gls{mua}.
		For both the slopes and iterative method, the accuracy of $\mu_s^{\prime,(\surd\rho)}$ depends on the true value of \gls{musp}, with lower true \glspl{musp} corresponding to worse accuracy.
		The purpose of these results is to indicate how accurate these methods are even in the case where the \glspl{rho} are known.
		From these results, we decided to focus on the iterative method due to its substantially better accuracy compared to the slopes method.
		\par
		
		\subsubsection{Error from Optode Position Displacement}
\begin{figure}[H]
			\includegraphics{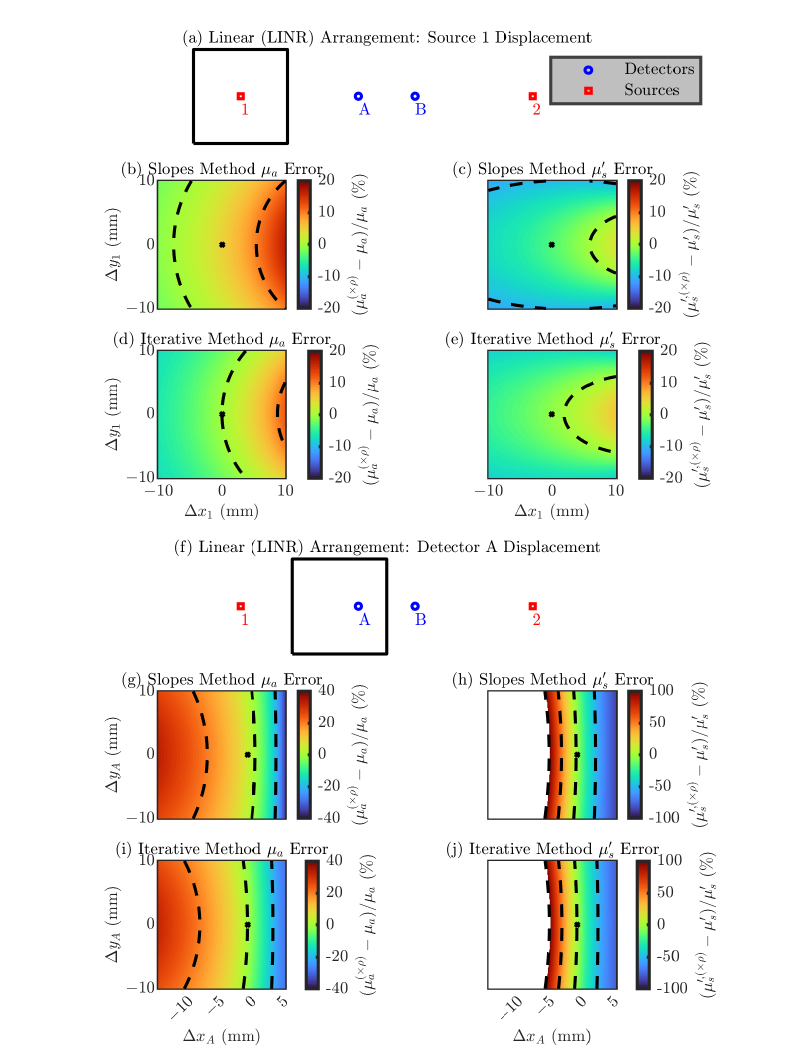}
			\caption{Error in the accuracy of the ``slopes'' or ``iterative'' optical recovery method for the LINeaR (LINR) arrangement.
				In the color-maps the black dashed iso-lines represent the color-bar tick-mark values and the nominal optode position is shown as a black dot.
				(a)--(e)~Errors from displacement of source 1 (equivalent to source 2).
				(f)--(j)~Errors from displacement of detector A (equivalent to detector B).
				(b)(c)(g)\&(h)~Results using the ``slopes'' \cite{Fantini_PMB99_NoninvasiveOptical} recovery method.
				(d)(e)(i)\&(j)~Results using the ``iterative'' \cite{Blaney_MDPIAppSci21_DualslopeDiffuse} recovery method.
				(b)(d)(g)\&(i)~Error in the \gls{mua}. 
				(c)(e)(h)\&(j)~Error in the \gls{musp}.
				\label{fig:LINRabs}\\
				\noindent{\footnotesize{Symbols: Displacement of optode from the nominal position ($[\Delta x, \Delta y]$), optical property recovered with the optode displaced from the nominal position ($(\times\rho)$ in superscript)}}
			}
		\end{figure}

		\paragraph{Single-Optode Displacement}
		\FloatBarrier
		\subparagraph{Linear (LINR) Arrangement}
		We begin investigating optode position errors by exploring errors in a single optode's position in the LINR arrangement.
		Due to the symmetry in this arrangement, displacing either source 1 or source 2 will result in the same error.
		Similarity, displacement in either detector A or detector B will also result in the same error.
		Therefore, we only show results for displacements of source 1 or detector A.
		\autoref{fig:LINRabs} shows the error in the recovery optical properties ($\mu_a^{(\times\rho)}$ and $\mu_s^{\prime,(\times\rho)}$) relative to the true optical properties (\as{mua} and \as{musp}) due to displacing the optode from its nominal position in a \SI{20x20}{\milli\meter} square.
		The recovered optical properties always assume the nominal optode position even if the optode is not in that position, thus the use of the $(\times\as{rho})$ superscript.
		\par
		
		\glsreset{mua}
		\glsreset{musp}
		
		To interpret the results in \autoref{fig:LINRabs} we will focus on the direction of maximum change in the error and the range of the error from displacement in the \SI{20x20}{\milli\meter} square.
		Concentrating on the subplots for source 1 (\autoref{fig:LINRabs}(a)--(e)) we find four color-maps, for each combination of the two recovery methods ``slopes'' or ``iterative'' (\autoref{fig:LINRabs}(b)--(c) or (d)--(e), respectively) and two optical properties \gls{mua} or \gls{musp} (\autoref{fig:LINRabs}(b)(d) or (c)(e), respectively).
		All four of these subplots show that larger errors occur if the optode is displaced away or toward the detectors (along the linear line of the LINR arrangement).
		In each case, the range of the error is about \SI{10}{\percent}.
		Now focusing on errors from displacement of detector A (\autoref{fig:LINRabs}(f)--(j)) we see again that movement along the linear line of the arrangement creates large changes in the error. 
		However, the magnitude of the error from displacing detector A is much large than source 1.
		Displacement of detector A in the \SI{20x20}{\milli\meter} square results in a error range of about \SI{30}{\percent} for \gls{mua} and \SI{>100}{\percent} for \gls{musp} (for both the slopes and iterative recovery methods).
		Overall, \autoref{fig:LINRabs} shows that displacement of optodes along the line of the LINR arrangement creates the largest error and the position of the detectors is much more critical than the position of the sources, but the two recovery methods do not have a significantly different dependence on optode position error.
		\par
		
		\autoref{tab:LINRabs} represents the magnitude of the error in the recovered optical properties from a \SI{1}{\milli\meter} displacement in any direction.
		This is done using the metrics ${\sigma^{\varnothing\SI{2}{\milli\meter}}_{\as{mua}}}/{\mu_a^{(\surd\rho)}}$ or ${\sigma^{\varnothing\SI{2}{\milli\meter}}_{\as{musp}}}/{\mu_s^{\prime,(\surd\rho)}}$ (\autoref{equ:RMSEmua} or \autoref{equ:RMSEmusp}) which represent the average error resulting from displacing an optode \SI{1}{\milli\meter} in any direction from its nominal position relative to the value the method would recover with the optode in the nominal position ($\mu_a^{(\surd\rho)}$ or $\mu_s^{\prime,(\surd\rho)}$).
		These values are important because they represent expected errors that could result from probe construction, since it is reasonable that optode positions in a real probe may be incorrect on the order of \SI{1}{\milli\meter}.
		Therefore, these errors can be thought of as expected systematic errors from the practice limitations of building a \gls{NIRS} probe, for a given arrangement, optical property, and recovery method.
		Specifically for the LINR arrangement, we see an error in the detector position propagates to a worse error in optical properties in all cases. 
		Additionally, the effect of a \SI{1}{\milli\meter} displacement does not seem significantly different between the slopes and iterative method.
		Aside from that, the remaining conclusions are somewhat expected from the limitations of diffusion theory.
		That is, overall larger errors in \gls{musp} than \gls{mua} (specifically for the case of detector displacement) and worse errors at low true values of \gls{mua} and \gls{musp}.
		For overall guidance, from a \SI{1}{\milli\meter} optode position error in any direction we can expect about
		a \SI{0.7}{\percent} or \SI{3}{\percent} error, for source or detector displacement, respectively, in \gls{mua} and about a 
		\SI{0.3}{\percent} or \SI{10}{\percent} error, for source or detector displacement, respectively, in \gls{musp} (using the iterative recovery method) for the LINR arrangement. 
		\par
		
		\begin{table}[H] 
			\caption{Errors in the recovered optical properties from a \SI{1}{\milli\meter} displacement in any direction relative to the value recovered in the nominal position, for the LINeR (LINR) arrangement \label{tab:LINRabs}}
			\begin{tabular}{
					S[table-format=1.3]S[table-format=1.1]|
					c|
					S[table-format=1.4]S[table-format=1.4]|
					S[table-format=1.4]S[table-format=1.4]
				}
				\toprule
				
				& 
				& 
				& 
				\multicolumn{4}{c}{\textbf{Recovery Method}} \\
				
				\textbf{\acrshort{mua}} & 
				\textbf{\acrshort{musp}} & 
				\textbf{Optode} &
				\multicolumn{2}{c|}{\textbf{Slopes}} &
				\multicolumn{2}{c}{\textbf{Iterative}} \\
				
				\textbf{(\si{\per\milli\meter})} & 
				\textbf{(\si{\per\milli\meter})} & 
				&
				\textbf{$\frac{\sigma^{\varnothing\SI{2}{\milli\meter}}_{\as{mua}}}{\mu_a^{(\surd\rho)}}$ (\si{\percent})} & 
				\textbf{$\frac{\sigma^{\varnothing\SI{2}{\milli\meter}}_{\as{musp}}}{\mu_s^{\prime,(\surd\rho)}}$ (\si{\percent})} & 
				\textbf{$\frac{\sigma^{\varnothing\SI{2}{\milli\meter}}_{\as{mua}}}{\mu_a^{(\surd\rho)}}$ (\si{\percent})} & 
				\textbf{$\frac{\sigma^{\varnothing\SI{2}{\milli\meter}}_{\as{musp}}}{\mu_s^{\prime,(\surd\rho)}}$ (\si{\percent})} \\
				
				\midrule
				0.005 & 0.5 & 1 or 2 & 1.1 & 0.56 & 1.2 & 0.44 \\ 
				0.005 & 0.5 & A or B & 5.1 & 16 & 5.0 & 13 \\ 
				0.005 & 1.0 & 1 or 2 & 0.82 & 0.52 & 0.92 & 0.50 \\ 
				0.005 & 1.0 & A or B & 3.5 & 15 & 4.6 & 14 \\ 
				0.005 & 1.5 & 1 or 2 & 0.68 & 0.46 & 0.75 & 0.45 \\ 
				0.005 & 1.5 & A or B & 2.9 & 14 & 3.8 & 14 \\ 
				0.010 & 0.5 & 1 or 2 & 0.68 & 0.37 & 0.71 & 0.29 \\ 
				0.010 & 0.5 & A or B & 3.4 & 15 & 3.0 & 13 \\ 
				0.010 & 1.0 & 1 or 2 & 0.53 & 0.37 & 0.56 & 0.35 \\ 
				0.010 & 1.0 & A or B & 2.3 & 14 & 2.8 & 13 \\ 
				0.010 & 1.5 & 1 or 2 & 0.44 & 0.33 & 0.46 & 0.32 \\ 
				0.010 & 1.5 & A or B & 1.9 & 14 & 2.3 & 13 \\ 
				0.015 & 0.5 & 1 or 2 & 0.53 & 0.27 & 0.54 & 0.22 \\ 
				0.015 & 0.5 & A or B & 2.8 & 15 & 2.3 & 12 \\ 
				0.015 & 1.0 & 1 or 2 & 0.42 & 0.29 & 0.44 & 0.28 \\ 
				0.015 & 1.0 & A or B & 1.9 & 14 & 2.2 & 13 \\ 
				0.015 & 1.5 & 1 or 2 & 0.34 & 0.27 & 0.36 & 0.26 \\ 
				0.015 & 1.5 & A or B & 1.5 & 13 & 1.8 & 13 \\ 
				\bottomrule
			\end{tabular}
			\\[1ex]
			\noindent{\footnotesize{Symbols: \Acrfull{mua}, \acrfull{musp},
					root-mean-squared-error from orbiting the optode position around the nominal optode position in a \SI{2}{\milli\meter} diameter circle ($\sigma^{\varnothing\SI{2}{\milli\meter}}$), optical property recovered with the optode in the nominal position ($(\surd\rho)$ in superscript)}}\\
		\end{table}

		\FloatBarrier
		\subparagraph{Asymmetric-Linear (ALIN) Arrangement}
		Second, we explore errors in a single optode's position in the ALIN arrangement.
		Due to the lack of symmetry in this arrangement, displacement of each optode must be looked at individually.
		\par
		
		\glsreset{mua}
		\glsreset{musp}
		\begin{figure}[H]
			\includegraphics{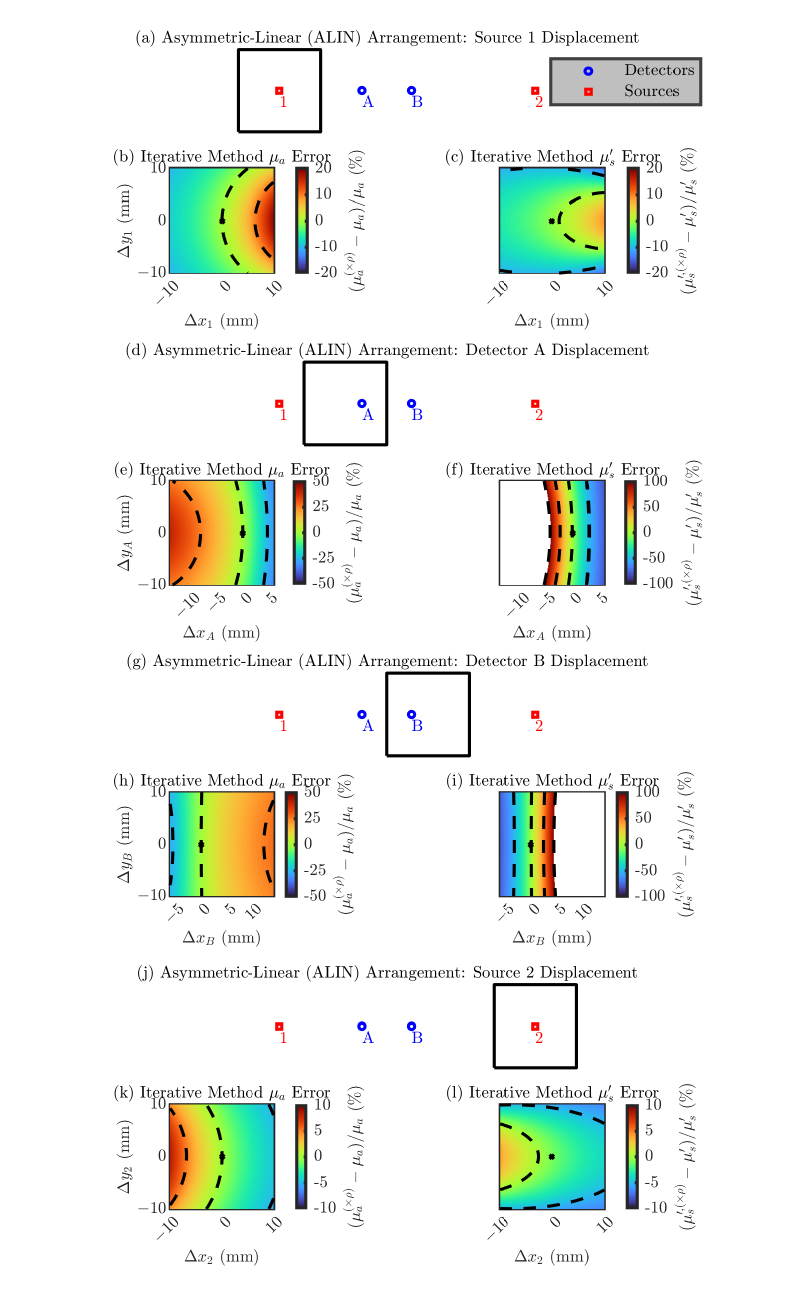}
			\caption{Error in the accuracy of the iterative optical recovery method for the Asymmetric-LINear (ALIN) arrangement.
				In the color-maps the black dashed iso-lines represent the color-bar tick-mark values and the nominal optode position is shown as a black dot.
				(a)--(c)~Errors from displacement of source 1.
				(d)--(f)~Errors from displacement of detector A.
				(g)--(i)~Errors from displacement of detector B.
				(j)--(l)~Errors from displacement of source 2.
				(b)(e)(h)\&(k)~Error in the \gls{mua}. 
				(c)(f)(i)\&(l)~Error in the \gls{musp}.
				\label{fig:ALINabs}\\
				\noindent{\footnotesize{Symbols: Displacement of optode from the nominal position ($[\Delta x, \Delta y]$), optical property recovered with the optode displaced from the nominal position ($(\times\rho)$ in superscript)}}
			}
		\end{figure}
		
		Interpreting the results in \autoref{fig:ALINabs}, we again focus on the direction of maximum change in the error and the range of the error in the simulated square.
		Similar to the LINR arrangement in \autoref{fig:LINRabs}, all of the subplots show that larger errors occur if the optode is displaced away or toward the detectors (along the linear line of the ALIN arrangement).
		Looking at the range of the error, we see that the position of the detectors is most critical with an error range of about \SI{60}{\percent} for \gls{mua} and \SI{>100}{\percent} for \gls{musp}.
		However, The position of source 1 is slightly more critical than the position of source 2.
		Source 1 shows an error range of about \SI{20}{\percent} for \gls{mua} and \gls{musp} while source 2 shows a range of about \SI{10}{\percent}.
		Overall, \autoref{fig:ALINabs} shows that displacement of optodes along the line of the ALIN arrangement creates the largest error and the position of the detectors is much more critical than the position of the sources, same as the LINR arrangement (\autoref{fig:LINRabs}).
		\par
		
		\autoref{tab:ALINabs} presents the magnitude of the error in the recovered optical properties from a \SI{1}{\milli\meter} displacement in any direction for the ALIN arrangement (\autoref{equ:RMSEmua} or \autoref{equ:RMSEmusp}).
		Similar to the LINR arrangement, we see for the ALIN arrangement errors in the detector position propagates to a worse error in optical properties in all cases. 
		To give overall guidance, from a \SI{1}{\milli\meter} optode position error in any direction we can expect about
		a \SI{0.7}{\percent} or \SI{3}{\percent} error, for source or detector displacement, respectively, in \gls{mua} and about a 
		\SI{0.4}{\percent} or \SI{10}{\percent} error, for source or detector displacement, respectively, in \gls{musp} (using the iterative recovery method) for the ALIN arrangement. 
		\par

		\begin{table}[H] 
			\caption{Errors in the recovered optical properties from a \SI{1}{\milli\meter} displacement in any direction relative to the value recovered in the nominal position, for the Asymmetric-LINear (ALIN) arrangement\label{tab:ALINabs}}
			\begin{tabular}{
					S[table-format=1.3]S[table-format=1.1]|
					c|
					S[table-format=1.4]S[table-format=1.4]
				}
				\toprule
				
				\textbf{\acrshort{mua}} & 
				\textbf{\acrshort{musp}} & 
				&
				&
				\\
				
				\textbf{(\si{\per\milli\meter})} & 
				\textbf{(\si{\per\milli\meter})} & 
				\textbf{Optode} &
				\textbf{$\frac{\sigma^{\varnothing\SI{2}{\milli\meter}}_{\as{mua}}}{\mu_a^{(\surd\rho)}}$ (\si{\percent})} & 
				\textbf{$\frac{\sigma^{\varnothing\SI{2}{\milli\meter}}_{\as{musp}}}{\mu_s^{\prime,(\surd\rho)}}$ (\si{\percent})} \\
				
				\midrule
				0.005 & 0.5 & 1 & 1.5 & 0.56 \\ 
				0.005 & 0.5 & 2 & 0.88 & 0.34 \\ 
				0.005 & 0.5 & A & 5.3 & 13 \\ 
				0.005 & 0.5 & B & 4.6 & 13 \\ 
				0.005 & 1.0 & 1 & 1.3 & 0.69 \\ 
				0.005 & 1.0 & 2 & 0.67 & 0.37 \\ 
				0.005 & 1.0 & A & 5.0 & 14 \\ 
				0.005 & 1.0 & B & 4.3 & 14 \\ 
				0.005 & 1.5 & 1 & 1.1 & 0.64 \\ 
				0.005 & 1.5 & 2 & 0.54 & 0.33 \\ 
				0.005 & 1.5 & A & 4.2 & 14 \\ 
				0.005 & 1.5 & B & 3.7 & 14 \\ 
				0.010 & 0.5 & 1 & 0.95 & 0.37 \\ 
				0.010 & 0.5 & 2 & 0.54 & 0.23 \\ 
				0.010 & 0.5 & A & 3.2 & 13 \\ 
				0.010 & 0.5 & B & 2.8 & 12 \\ 
				0.010 & 1.0 & 1 & 0.81 & 0.48 \\ 
				0.010 & 1.0 & 2 & 0.41 & 0.26 \\ 
				0.010 & 1.0 & A & 3.1 & 13 \\ 
				0.010 & 1.0 & B & 2.7 & 13 \\ 
				0.010 & 1.5 & 1 & 0.68 & 0.45 \\ 
				0.010 & 1.5 & 2 & 0.34 & 0.24 \\ 
				0.010 & 1.5 & A & 2.6 & 14 \\ 
				0.010 & 1.5 & B & 2.3 & 13 \\ 
				0.015 & 0.5 & 1 & 0.73 & 0.28 \\ 
				0.015 & 0.5 & 2 & 0.42 & 0.18 \\ 
				0.015 & 0.5 & A & 2.4 & 12 \\ 
				0.015 & 0.5 & B & 2.1 & 12 \\ 
				0.015 & 1.0 & 1 & 0.63 & 0.38 \\ 
				0.015 & 1.0 & 2 & 0.32 & 0.21 \\ 
				0.015 & 1.0 & A & 2.4 & 13 \\ 
				0.015 & 1.0 & B & 2.1 & 13 \\ 
				0.015 & 1.5 & 1 & 0.52 & 0.37 \\ 
				0.015 & 1.5 & 2 & 0.26 & 0.19 \\ 
				0.015 & 1.5 & A & 2.0 & 13 \\ 
				0.015 & 1.5 & B & 1.7 & 13 \\ 
				\bottomrule
			\end{tabular}
			\\[1ex]
			\noindent{\footnotesize{Symbols: \Acrfull{mua}, \acrfull{musp},
					root-mean-squared-error from orbiting the optode position around the nominal optode position in a \SI{2}{\milli\meter} diameter circle ($\sigma^{\varnothing\SI{2}{\milli\meter}}$), optical property recovered with the optode in the nominal position ($(\surd\rho)$ in superscript)}}\\
		\end{table}

		\FloatBarrier
		\subparagraph{Trapezoidal (TRAP) Arrangement}
		Next, we explore errors in a single optode's position in the TRAP arrangement.
		The symmetry of the TRAP arrangement lets us focus on only source 1 (since it is equivalent to source 2) and detector A (since it is equivalent to detector B).
		\par
		
		\glsreset{mua}
		\glsreset{musp}
		\begin{figure}[H]
			\includegraphics{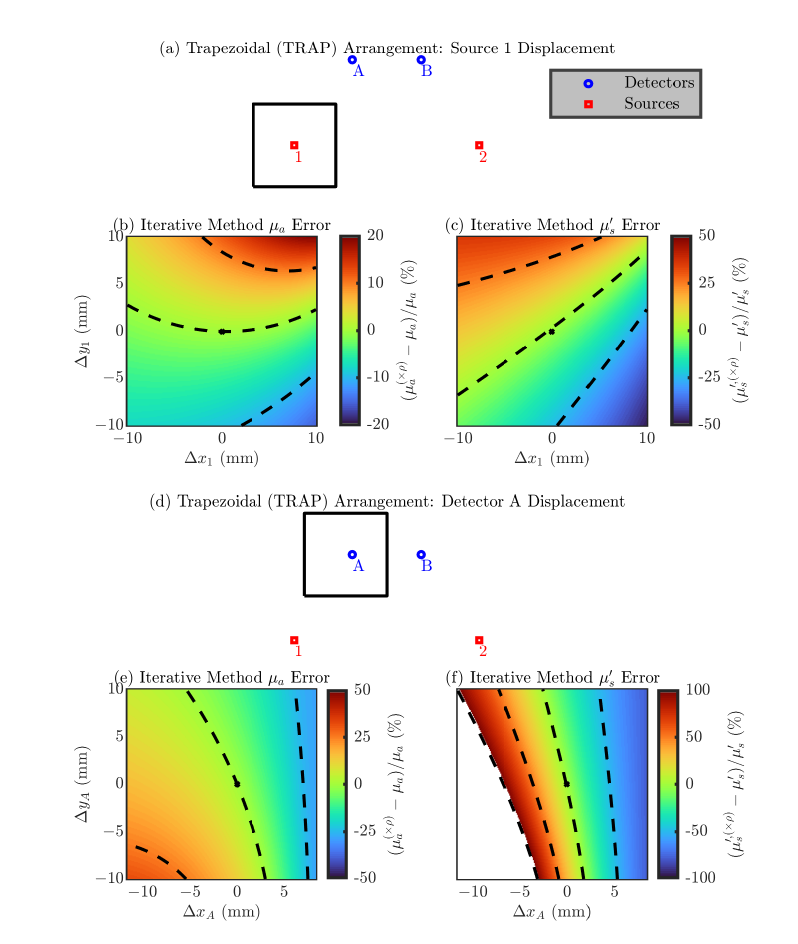}
			\caption{Error in the accuracy of the iterative optical recovery method for the TRAPezoidal (TRAP) arrangement.
				In the color-maps the black dashed iso-lines represent the color-bar tick-mark values and the nominal optode position is shown as a black dot.
				(a)--(c)~Errors from displacement of source 1 (equivalent to source 2).
				(d)--(f)~Errors from displacement of detector A (equivalent to detector B).
				(b)\&(e)~Error in the \gls{mua}. 
				(c)\&(f)~Error in the \gls{musp}.
				\label{fig:TRAPabs}\\
				\noindent{\footnotesize{Symbols: Displacement of optode from the nominal position ($[\Delta x, \Delta y]$), optical property recovered with the optode displaced from the nominal position ($(\times\rho)$ in superscript)}}
			}
		\end{figure}
		
		Focusing on the results in \autoref{fig:TRAPabs}, we again emphasize the direction of maximum change in the error and the range of the error in the simulated square.
		In this arrangement the directions which cause a maximum change in the error are not as trivial as they were in the LINR and ALIN arrangements.
		Looking at the range of the error we again see that the position of the detectors is most critical with an error range for detector A of about \SI{60}{\percent} for \gls{mua} and \SI{>100}{\percent} for \gls{musp}.
		Instead, source 1 shows an error range of about \SI{30}{\percent} for \gls{mua} and about \SI{70}{\percent} for \gls{musp}.
		Overall, \autoref{fig:TRAPabs} shows that the direction of the optode displacement which causes the maximum error is not as trivial as it is with the LINR or ALIN arrangements.
		We will discuss a way to use knowledge of these directions in \autoref{sec:DISoptSupp}.
		\par
		
		\autoref{tab:TRAPabs} shows the magnitude of the error in the recovered optical properties from a \SI{1}{\milli\meter} displacement in any direction for the TRAP arrangement (\autoref{equ:RMSEmua} or \autoref{equ:RMSEmusp}).
		Similar to the LINR and ALIN arrangement, we see for the TRAP arrangement errors in the detector position propagate to a worse error in optical properties in all cases, however, the difference in the dependence on source or detector is not as strong with the TRAP arrangement. 
		To give overall guidance, from a \SI{1}{\milli\meter} optode position error in any direction we can expect about
		a \SI{0.9}{\percent} or \SI{2}{\percent} error, for source or detector displacement, respectively, in \gls{mua} and about a 
		\SI{2}{\percent} or \SI{9}{\percent} error, for source or detector displacement, respectively, in \gls{musp} (using the iterative recovery method) for the TRAP arrangement. 
		\par
		
		\begin{table}[H] 
			\caption{Errors in the recovered optical properties from a \SI{1}{\milli\meter} displacement in any direction relative to the value recovered in the nominal position, for the TRAPezoidal (TRAP) arrangement\label{tab:TRAPabs}}
			\begin{tabular}{
					S[table-format=1.3]S[table-format=1.1]|
					c|
					S[table-format=1.4]S[table-format=1.4]
				}
				\toprule
				
				\textbf{\acrshort{mua}} & 
				\textbf{\acrshort{musp}} & 
				&
				&
				\\
				
				\textbf{(\si{\per\milli\meter})} & 
				\textbf{(\si{\per\milli\meter})} & 
				\textbf{Optode} &
				\textbf{$\frac{\sigma^{\varnothing\SI{2}{\milli\meter}}_{\as{mua}}}{\mu_a^{(\surd\rho)}}$ (\si{\percent})} & 
				\textbf{$\frac{\sigma^{\varnothing\SI{2}{\milli\meter}}_{\as{musp}}}{\mu_s^{\prime,(\surd\rho)}}$ (\si{\percent})} \\
				
				\midrule
				0.005 & 0.5 & 1 or 2 & 1.5 & 2.5 \\ 
				0.005 & 0.5 & A or B & 3.7 & 9.3 \\ 
				0.005 & 1.0 & 1 or 2 & 1.2 & 2.7 \\ 
				0.005 & 1.0 & A or B & 3.3 & 9.9 \\ 
				0.005 & 1.5 & 1 or 2 & 1.0 & 2.7 \\ 
				0.005 & 1.5 & A or B & 2.7 & 9.9 \\ 
				0.010 & 0.5 & 1 or 2 & 0.91 & 2.4 \\ 
				0.010 & 0.5 & A or B & 2.2 & 8.9 \\ 
				0.010 & 1.0 & 1 or 2 & 0.77 & 2.6 \\ 
				0.010 & 1.0 & A or B & 2.0 & 9.5 \\ 
				0.010 & 1.5 & 1 or 2 & 0.63 & 2.6 \\ 
				0.010 & 1.5 & A or B & 1.7 & 9.5 \\ 
				0.015 & 0.5 & 1 or 2 & 0.69 & 2.4 \\ 
				0.015 & 0.5 & A or B & 1.7 & 8.8 \\ 
				0.015 & 1.0 & 1 or 2 & 0.59 & 2.5 \\ 
				0.015 & 1.0 & A or B & 1.6 & 9.3 \\ 
				0.015 & 1.5 & 1 or 2 & 0.49 & 2.5 \\ 
				0.015 & 1.5 & A or B & 1.3 & 9.3 \\ 
				\bottomrule
			\end{tabular}
			\\[1ex]
			\noindent{\footnotesize{Symbols: \Acrfull{mua}, \acrfull{musp},
					root-mean-squared-error from orbiting the optode position around the nominal optode position in a \SI{2}{\milli\meter} diameter circle ($\sigma^{\varnothing\SI{2}{\milli\meter}}$), optical property recovered with the optode in the nominal position ($(\surd\rho)$ in superscript)}}
		\end{table}

		\FloatBarrier
		\subparagraph{Diagonal-Rectangular (DRCT) Arrangement}
		Finally, we explore errors in a single optode's position in the DRCT arrangement.
		The symmetry of the DRCT arrangement lets us focus on only source 1 (since it is equivalent to source 2, detector A, and detector B).
		\par
		
		\glsreset{mua}
		\glsreset{musp}
		\begin{figure}[H]
			\includegraphics{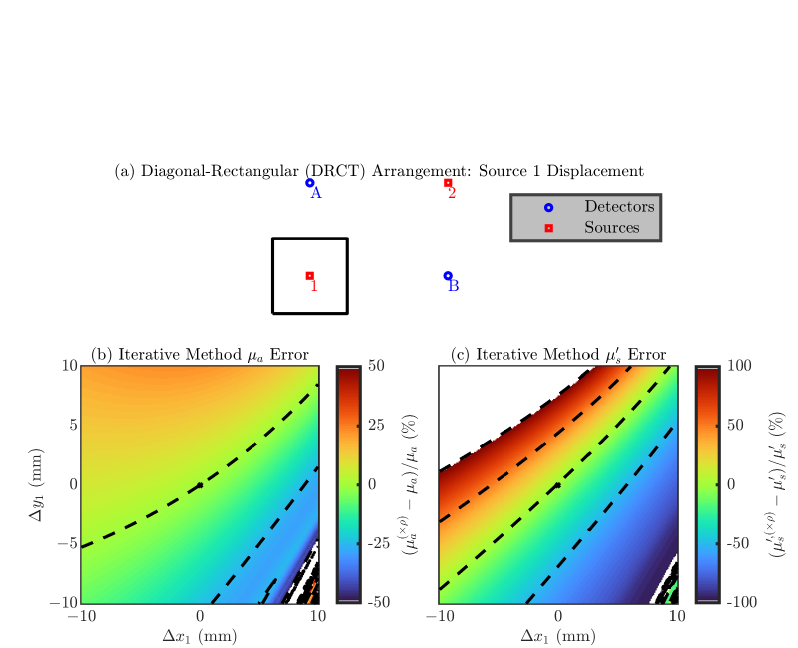}
			\caption{Error in the accuracy of the iterative optical recovery method for the Diagonal-ReCTangular (DRCT) arrangement.
				In the color-maps the black dashed iso-lines represent the color-bar tick-mark values and the nominal optode position is shown as a black dot.
				(a)--(c)~Errors from displacement of source 1 (equivalent to source 2, detector A, and detector B).
				(b)~Error in the \gls{mua}. 
				(c)~Error in the \gls{musp}.
				\label{fig:DRCTabs}\\
				\noindent{\footnotesize{Symbols: Displacement of optode from the nominal position ($[\Delta x, \Delta y]$), optical property recovered with the optode displaced from the nominal position ($(\times\rho)$ in superscript)}}
			}
		\end{figure}
		
		\autoref{fig:DRCTabs} shows the results for the DRCT arrangement.
		Similar to the TRAP arrangement, the directions which cause a maximum change in the error are not as trivial as they were in the LINR and ALIN arrangements.
		However, roughly speaking, it seems that displacement of an optode towards or away from the center causes little error (i.e., movement orthogonal to that creates a large change in the error).
		Since for this arrangement displacement of any optode is equivalent, there is no error dependence preference between sources and detectors (i.e., there positions are equally important).
		Looking at the range of the error we see a error range of about \SI{>100}{\percent} for \gls{mua} and \SI{>100}{\percent} for \gls{musp} from moving with the \SI{20x20}{\milli\meter} square.
		However, the large (i.e., \SI{>100}{\percent}) errors occur for \gls{mua} only in the lower right corner of the square, ignoring that region the \gls{mua} error range is about \SI{50}{\percent}.
		\par
		
		\autoref{tab:DRCTabs} shows the magnitude of the error in the recovered optical properties from a \SI{1}{\milli\meter} displacement in any direction for the DRCT arrangement (\autoref{equ:RMSEmua} or \autoref{equ:RMSEmusp}).
		In this case there is no dependence on optode, since displacement of any optode is equivalent.
		To give overall guidance, from a \SI{1}{\milli\meter} optode position error in any direction we can expect about
		a \SI{2}{\percent} in \gls{mua} and about a 
		\SI{9}{\percent} in \gls{musp} (using the iterative recovery method) for the DRCT arrangement. 
		\par
		
		\begin{table}[H] 
			\caption{Errors in the recovered optical properties from a \SI{1}{\milli\meter} displacement in any direction relative to the value recovered in the nominal position, for the Diagonal-ReCTangular (DRCT) arrangement\label{tab:DRCTabs}}
			\begin{tabular}{
					S[table-format=1.3]S[table-format=1.1]|
					c|
					S[table-format=1.4]S[table-format=1.4]
				}
				\toprule
				
				\textbf{\acrshort{mua}} & 
				\textbf{\acrshort{musp}} & 
				&
				&
				\\
				
				\textbf{(\si{\per\milli\meter})} & 
				\textbf{(\si{\per\milli\meter})} & 
				\textbf{Optode} &
				\textbf{$\frac{\sigma^{\varnothing\SI{2}{\milli\meter}}_{\as{mua}}}{\mu_a^{(\surd\rho)}}$ (\si{\percent})} & 
				\textbf{$\frac{\sigma^{\varnothing\SI{2}{\milli\meter}}_{\as{musp}}}{\mu_s^{\prime,(\surd\rho)}}$ (\si{\percent})} \\
				
				\midrule
				0.005 & 0.5 & 1, 2, A, or B & 3.7 & 9.3 \\ 
				0.005 & 1.0 & 1, 2, A, or B & 3.3 & 9.9 \\ 
				0.005 & 1.5 & 1, 2, A, or B & 2.7 & 9.9 \\ 
				0.010 & 0.5 & 1, 2, A, or B & 2.2 & 8.9 \\ 
				0.010 & 1.0 & 1, 2, A, or B & 2.0 & 9.5 \\ 
				0.010 & 1.5 & 1, 2, A, or B & 1.7 & 9.5 \\ 
				0.015 & 0.5 & 1, 2, A, or B & 1.7 & 8.8 \\ 
				0.015 & 1.0 & 1, 2, A, or B & 1.6 & 9.3 \\ 
				0.015 & 1.5 & 1, 2, A, or B & 1.3 & 9.3 \\ 
				\bottomrule
			\end{tabular}
			\\[1ex]
			\noindent{\footnotesize{Symbols: \Acrfull{mua}, \acrfull{musp},
					root-mean-squared-error from orbiting the optode position around the nominal optode position in a \SI{2}{\milli\meter} diameter circle ($\sigma^{\varnothing\SI{2}{\milli\meter}}$), optical property recovered with the optode in the nominal position ($(\surd\rho)$ in superscript)}}
		\end{table}

		\paragraph{Multi-Optode Displacement}
		\autoref{tab:absMulti} shows the average errors which may be expected if multiple optodes are displaced by \SI{1}{\milli\meter} in any direction assuming \SI{1.0}{\per\milli\meter} for \gls{musp} and \SI{0.01}{\per\milli\meter} for \gls{mua}.
		Certain combinations were omitted due to symmetry as follows (the ``$\equiv$'' symbol is used to indicate that an optode combination is equivalent): 
		for LINR and TRAP 1A~$\equiv$~2B, 1B~$\equiv$~2A, 1AB~$\equiv$~2AB, and 12A~$\equiv$~12B;
		for ALIN no combination is omitted;
		and for DRCT 1A~$\equiv$~2B, 1B~$\equiv$~2A, 12~$\equiv$~AB, and 1AB~$\equiv$~2AB~$\equiv$~12A~$\equiv$~12B.
		The errors in this table are, for the most part, no more than twice the worst case error from a single optode displacement.
		Furthermore, moving from two optodes to three to four does not significantly change the magnitude of the error, with more dependence on if a critical optode is displaced (e.g., the detectors in the LINR arrangement).
		For example in the LINR case, the lowest error occurs by displacing 12 which is the only case that does not include a detector displacement.
		Furthermore, in this case 12A shows an error more compatible to the 1A and 1B cases then the 1AB case, therefore the main influence in the magnitude of the error appears to be how many detectors as displaced (i.e., more generally, how many critical optodes are displaced).
		Overall, these results point to the TRAP arrangement being slightly more robust against optode displacement than the other arrangements, possibly because it de-emphasizes the criticality of the detector positions.
		This can be seen in the case were all four optodes are displaced (12AB) which results in TRAP average errors of about \SI{3}{\percent} and \SI{14}{\percent}, for \gls{mua} and \gls{musp}, respectively.
		While, this average error is \SI{4}{\percent} and \SI{19}{\percent}, for \gls{mua} and \gls{musp}, respectively, for the LINR, ALIN, and DRCT arrangements.
		\par

		\begin{table}[H] 
			\caption{Errors in the recovered optical properties from a simultaneous \SI{1}{\milli\meter} displacement by multiple optodes in any direction relative to the value recovered in the nominal positions\label{tab:absMulti}}
			\begin{tabular}{
					c|
					c|
					S[table-format=1.4]S[table-format=1.4]
				}
				\toprule
				
				\textbf{Arrangement} & 
				\textbf{Optodes} &
				\textbf{$\frac{\sigma^{\varnothing\SI{2}{\milli\meter}}_{\as{mua}}}{\mu_a^{(\surd\rho)}}$ (\si{\percent})} & 
				\textbf{$\frac{\sigma^{\varnothing\SI{2}{\milli\meter}}_{\as{musp}}}{\mu_s^{\prime,(\surd\rho)}}$ (\si{\percent})} \\
				
				\midrule
				\multirow{7}{*}{\rotatebox[origin=c]{90}{LINR}} &
				1 and A & 2.9 & 13 \\ 
				& 1 and B & 2.9 & 13 \\ 
				& 1 and 2 & 0.80 & 0.49 \\ 
				& A and B & 4.0 & 19 \\ 
				& 1, A, and B & 4.0 & 19 \\ 
				& 1, 2, and A & 2.9 & 13 \\ 
				& 1, 2, A, and B & 4.1 & 19 \\ 
				\midrule
				\multirow{11}{*}{\rotatebox[origin=c]{90}{ALIN}} &
				1 and A & 3.2 & 13 \\ 
				& 1 and B & 2.8 & 13 \\ 
				& 2 and A & 3.1 & 13 \\ 
				& 2 and B & 2.7 & 13 \\ 
				& 1 and 2 & 0.91 & 0.55 \\ 
				& A and B & 4.1 & 19 \\ 
				& 1, A, and B & 4.2 & 19 \\ 
				& 2, A, and B & 4.1 & 19 \\ 
				& 1, 2, and A & 3.2 & 13 \\ 
				& 1, 2, and B & 2.8 & 13 \\ 
				& 1, 2, A, and B & 4.2 & 19 \\ 
				\midrule
				\multirow{7}{*}{\rotatebox[origin=c]{90}{TRAP}} &
				1 and A & 2.2 & 9.8 \\ 
				& 1 and B & 2.2 & 9.8 \\ 
				& 1 and 2 & 1.1 & 3.6 \\ 
				& A and B & 2.9 & 13 \\ 
				& 1, A, and B & 3.0 & 14 \\ 
				& 1, 2, and A & 2.3 & 10 \\ 
				& 1, 2, A, and B & 3.1 & 14 \\ 
				\midrule
				\multirow{5}{*}{\rotatebox[origin=c]{90}{DRCT}} & 
				1 and A & 2.9 & 13 \\ 
				& 1 and B & 2.9 & 13 \\ 
				& 1 and 2 & 2.9 & 13 \\ 
				& 1, A, and B & 3.5 & 16 \\ 
				& 1, 2, A, and B & 4.1 & 19 \\ 
				\bottomrule
			\end{tabular}
			\\[1ex]
			\noindent{\footnotesize{Symbols: \Acrfull{mua}, \acrfull{musp},
					root-mean-squared-error from orbiting the optode position around the nominal optode position in a \SI{2}{\milli\meter} diameter circle ($\sigma^{\varnothing\SI{2}{\milli\meter}}$), optical property recovered with the optode in the nominal position ($(\surd\rho)$ in superscript)}}
		\end{table}

		\subsection{Dual-Slope Recovery of Relative Changes in the \Acrfull{mua}}\label{sec:RESds}
\begin{figure}[H]
			\includegraphics{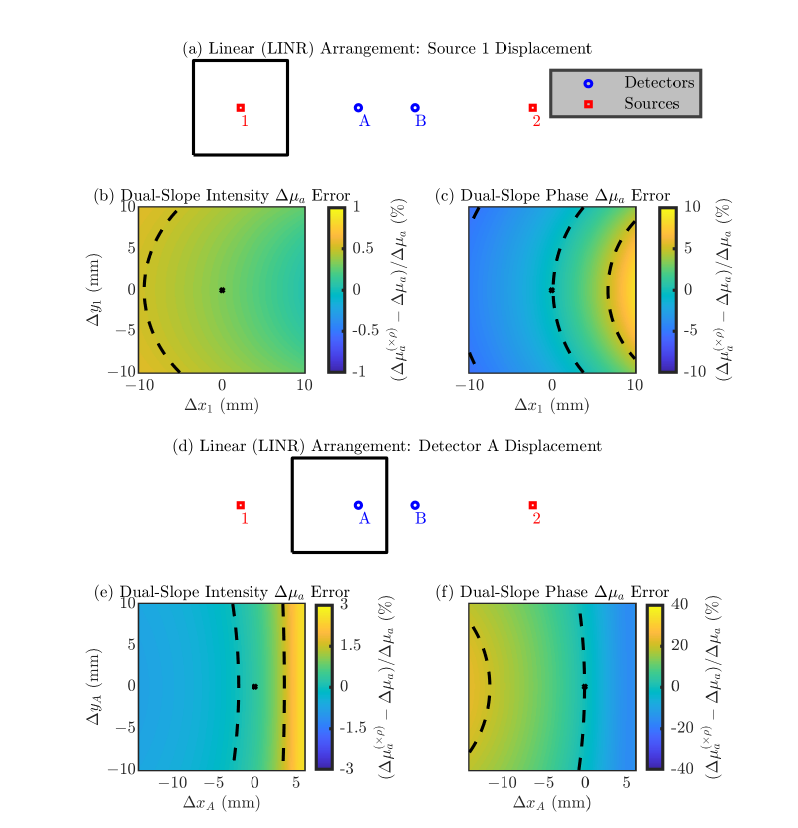}
			\caption{Error in the accuracy of recovering \gls{dmua} for the LINeaR (LINR) arrangement.
				In the color-maps the black dashed iso-lines represent the color-bar tick-mark values and the nominal optode position is shown as a black dot.
				(a)--(c)~Errors from displacement of source 1 (equivalent to source 2).
				(d)--(f)~Errors from displacement of detector A (equivalent to detector B).
				(b)\&(e)~Error in the \gls{dmua} recovered by \gls{DS} \gls{I}. 
				(c)\&(f)~Error in the \gls{dmua} recovered by \gls{DS} \gls{phi}. 
				\label{fig:LINRrel}\\
				\noindent{\footnotesize{Symbols: Displacement of optode from the nominal position ($[\Delta x, \Delta y]$), optical property recovered with the optode displaced from the nominal position ($(\times\rho)$ in superscript)}}
			}
		\end{figure}

\begin{figure}[H]
			\includegraphics{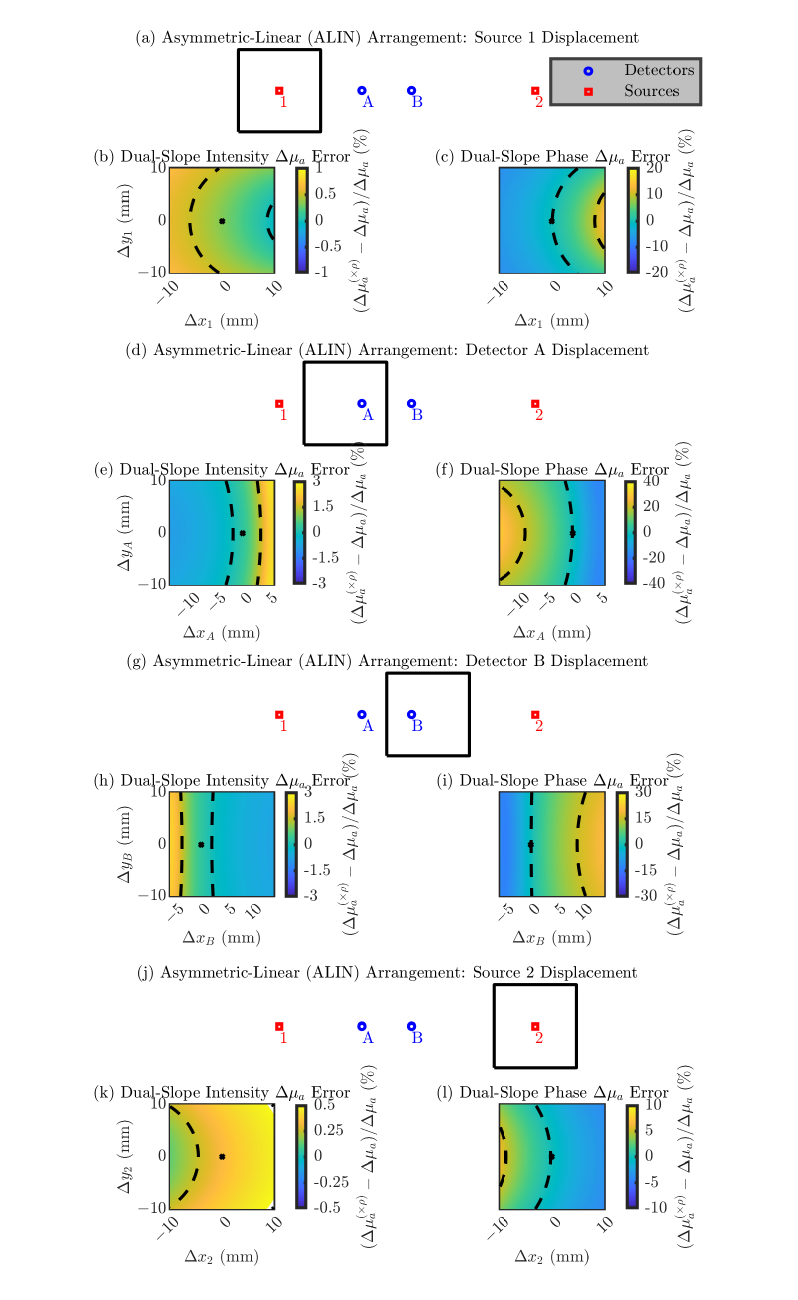}
			\caption{Error in the accuracy of recovering \gls{dmua} for the Asymmetric-LINear (ALIN) arrangement.
				In the color-maps the black dashed iso-lines represent the color-bar tick-mark values and the nominal optode position is shown as a black dot.
				(a)--(c)~Errors from displacement of source 1.
				(d)--(f)~Errors from displacement of detector A.
				(g)--(i)~Errors from displacement of detector B.
				(j)--(l)~Errors from displacement of source 2.
				(b)(e)(h)\&(k)~Error in the \gls{dmua} recovered by \gls{DS} \gls{I}. 
				(c)(f)(i)\&(l)~Error in the \gls{dmua} recovered by \gls{DS} \gls{phi}. 
				\label{fig:ALINrel}\\
				\noindent{\footnotesize{Symbols: Displacement of optode from the nominal position ($[\Delta x, \Delta y]$), optical property recovered with the optode displaced from the nominal position ($(\times\rho)$ in superscript)}}
			}
		\end{figure}

\begin{figure}[H]
			\includegraphics{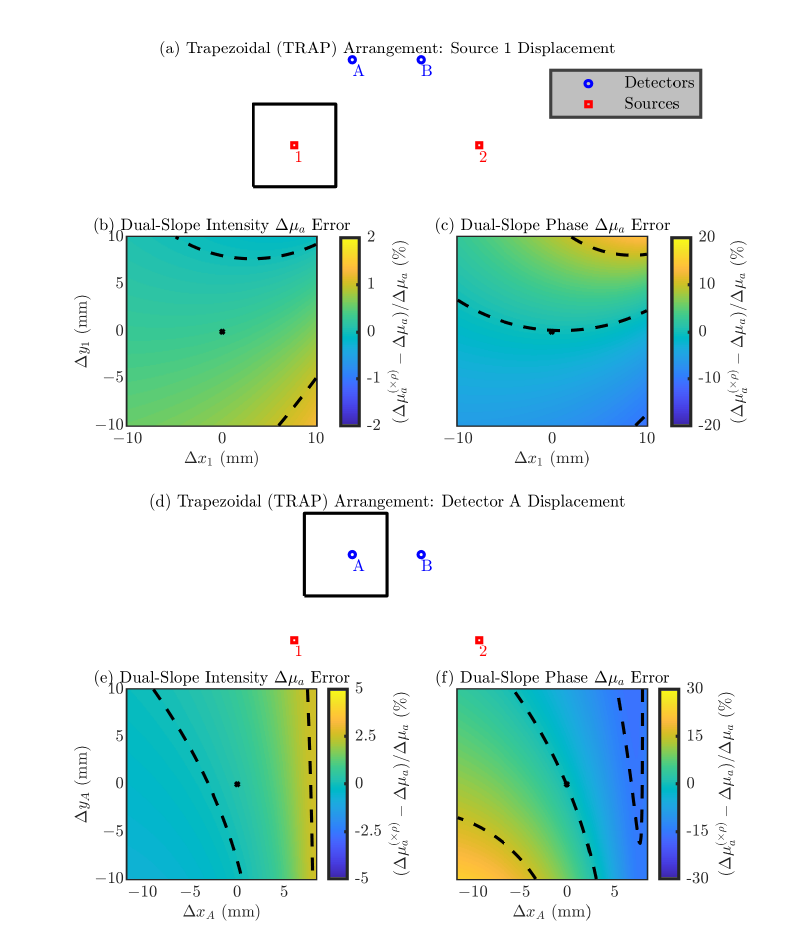}
			\caption{Error in the accuracy of recovering \gls{dmua} for the TRAPezoidal (TRAP) arrangement.
				In the color-maps the black dashed iso-lines represent the color-bar tick-mark values and the nominal optode position is shown as a black dot.
				(a)--(c)~Errors from displacement of source 1 (equivalent to source 2).
				(d)--(f)~Errors from displacement of detector A (equivalent to detector B).
				(b)\&(e)~Error in the \gls{dmua} recovered by \gls{DS} \gls{I}. 
				(c)\&(f)~Error in the \gls{dmua} recovered by \gls{DS} \gls{phi}. 
				\label{fig:TRAPrel}\\
				\noindent{\footnotesize{Symbols: Displacement of optode from the nominal position ($[\Delta x, \Delta y]$), optical property recovered with the optode displaced from the nominal position ($(\times\rho)$ in superscript)}}
			}
		\end{figure}

\begin{figure}[H]
			\includegraphics{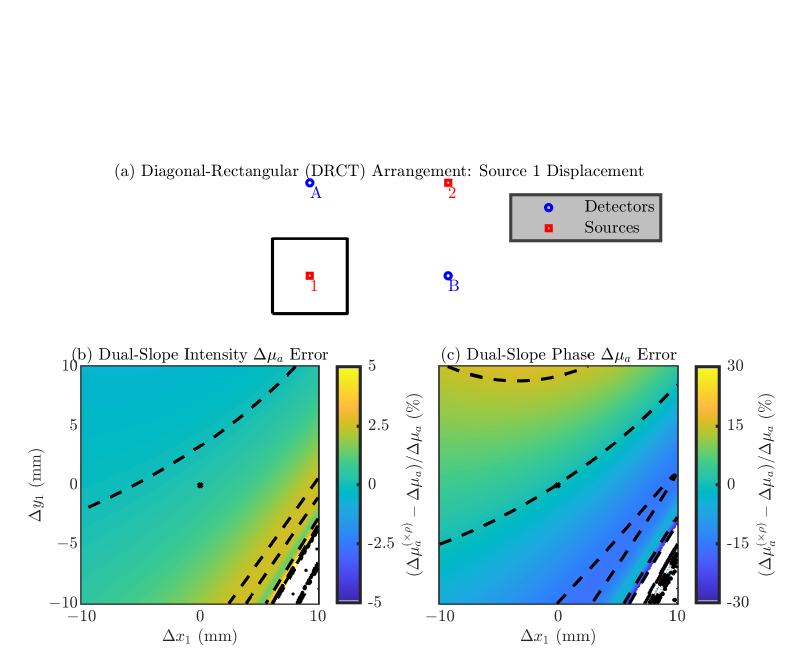}
			\caption{Error in the accuracy of recovering \gls{dmua} for the Diagonal-ReCTangular (DRCT) arrangement.
				In the color-maps the black dashed iso-lines represent the color-bar tick-mark values and the nominal optode position is shown as a black dot.
				(a)--(c)~Errors from displacement of source 1 (equivalent to source 2, detector A, and detector B).
				(b)~Error in the \gls{dmua} recovered by \gls{DS} \gls{I}. 
				(c)~Error in the \gls{dmua} recovered by \gls{DS} \gls{phi}. 
				\label{fig:DRCTrel}\\
				\noindent{\footnotesize{Symbols: Displacement of optode from the nominal position ($[\Delta x, \Delta y]$), optical property recovered with the optode displaced from the nominal position ($(\times\rho)$ in superscript)}}
			}
		\end{figure}

		In the following section the figures are in the same format as \autoref{sec:RESsc} but show results relevant to \gls{DS} measurements of \gls{dmua} instead of \gls{SC} measurements of the absolute \gls{mua} \& \gls{musp}.
		These results show errors in the recovered \gls{dmua} (shown with a $(\times\as{rho})$ in superscript to indicate that the nominal position is always assumed) relative to the true \gls{dmua} simulated in the medium.
		Using \gls{DS}, \gls{dmua} can be recovered with either only \gls{I} data or only \gls{phi} data, therefore results are shown for both cases.
		\autoref{fig:LINRrel}, \autoref{fig:ALINrel}, \autoref{fig:TRAPrel}, and \autoref{fig:DRCTrel}
		show these results for,
		the LINR, the ALIN, the TRAP, and the DRCT arrangements, respectively.
		As before, we suggest one focuses on the direction of the change in the error and the magnitude of the range in the error from displacing an optode in a \SI{20x20}{\milli\meter} square around its nominal position.
		\par
		
		Summarizing these results, we note that, overall, the relative errors in \gls{dmua} due to optode position errors are less then those for absolute optical properties.
		For \gls{DS} \gls{I} the error is often on the order of \SI{1}{\percent} from moving the optode in the \SI{20x20}{\milli\meter} square.
		However, for \gls{DS} \gls{phi} the order is \SI{10}{\percent}.
		Therefore, the ability for \gls{DS} \gls{phi} to accurately recover values of \gls{dmua} depends much more on the optode position then \gls{DS} \gls{I} by an order of magnitude.
		\par
		
		\glsreset{dmua}
		\glsreset{DS}
		\glsreset{I}
		\glsreset{phi}
		
		\glsreset{dmua}
		\glsreset{DS}
		\glsreset{I}
		\glsreset{phi}
		
		\glsreset{dmua}
		\glsreset{DS}
		\glsreset{I}
		\glsreset{phi}
		
		\glsreset{dmua}
		\glsreset{DS}
		\glsreset{I}
		\glsreset{phi}

		%%%%%%%%%%%%%%%%%%%%%%%%%%%%%%%%%%%%%%%%%%
		\section{Discussion}
		Based on the results in this work, we can provide guidance on what type of \gls{SC} / \gls{DS} arrangement is more robust against optode position errors.
		Additionally, we can provide guidance on what absolute optical property recovery method to use.
		Regarding the latter, we advocate for the iterative absolute optical property recovery method \cite{Blaney_MDPIAppSci21_DualslopeDiffuse} on \gls{FD} \gls{SC} data.
		We justify this by comparing the accuracy errors of the iterative and slopes method in a case where the optode positions are assumed correctly (\autoref{tab:absMethAcc}).
		Focusing on the error in the recovered \gls{mua}, the iterative method showed an average error of \SI{6}{\percent} while the iterative method showed \SI{0.01}{\percent}; with a similar relationship for the error in \gls{musp}.
		Additionally, if we consider how robust these two methods are against optode position errors (\autoref{tab:LINRabs}), we see that the iterative method either performs the same as the slopes method or has a slight advantage.
		Therefore, we conclude the iterative preferable over the slopes method.
		\par
		
		For a comparison of the \gls{SC} / \gls{DS} arrangements, we can examine the results in both \autoref{sec:RESsc} and \autoref{sec:RESds}.
		Considering all of these together, we conclude that arrangements typically have a critical optode type, and movement of the critical optode causes large errors in the recovered values.
		The critical optode type is characterized by the type of optodes placed near the arrangement's line of symmetry or centroid (e.g., detectors A and B in LINR).
		The critical type may be either sources or detectors since sources and detectors may be interchanged (e.g., if sources and detectors are switched in LINR then sources 1 and 2 are critical).
		For the LINR and ALIN the critical optodes are optodes A \& B, while for TRAP the importance of the optodes 1 \& 2 and optodes A \& B is more equally balanced but optodes A \& B are still slightly more critical then the optodes 1 \& 2.
		The DRCT arrangement is not ideal since all the optode's positions are critical since displacement of any optode is geometrically equivalent to the displacement of another.
		Additionally, the results in \autoref{tab:absMulti} further suggest that the TRAP arrangement may be more robust against multiple optode position errors then the other arrangements.
		A second conclusion is that the error in \gls{musp} or \gls{dmua} recovered by \gls{DS} \gls{phi} is more dependent on the position of the critical optodes then \gls{mua} or \gls{dmua} recovered by \gls{DS} \gls{I}, respectively.
		\par
		
		Results in \autoref{sec:RESsc} unsurprisingly show that \gls{musp} is more susceptible to errors, but from the recovery method itself and from optode position errors.
		However, a more unexpected result is that from \autoref{sec:RESds}, which shows that \gls{dmua} recovered by \gls{DS} \gls{phi} is more susceptible to errors then \gls{DS} \gls{I} by an order of magnitude.
		This may have implications when comparing \glspl{dmua} recovered by either \gls{DS} \gls{I} or \gls{DS} \gls{phi} since a small optode position error could propagate to almost no error in \gls{DS} \gls{I}'s results but a large error in \gls{DS} \gls{phi}'s results, making comparison of the data-types tenuous.
		\par
		
		\subsection{Design of Optode Support Structure}\label{sec:DISoptSupp}
\begin{figure}[H]
			\includegraphics{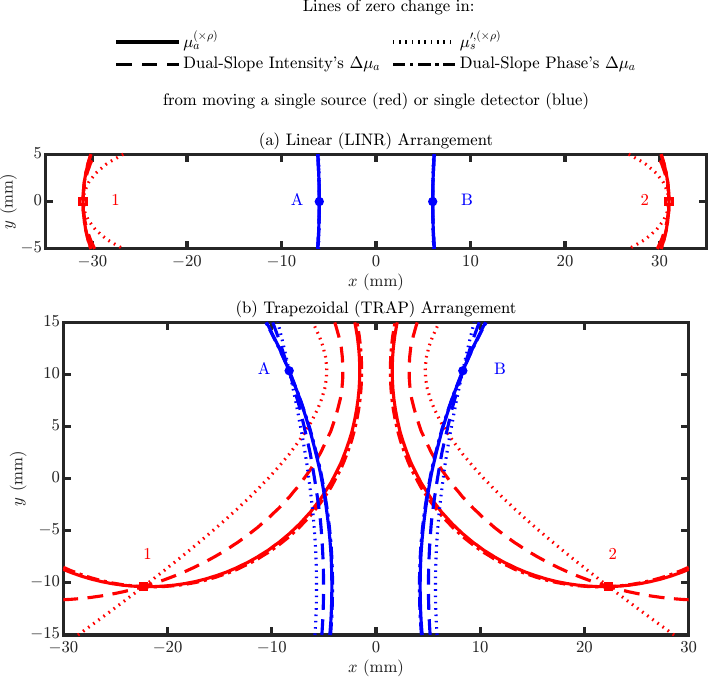}
			\caption{Lines for which if the optode is moved along them a recovered optical property will not change (i.e., lines of zero change). E.g., the solid red line which passes through source 1 shows positions where source 1 can be placed and the \gls{mua} recovered by assuming the nominal \glspl{rho} will be the same.
				(a) The LINeaR (LINR) arrangement.
				(b) The TRAPezoidal (TRAP) arrangement.
				\label{fig:linesNoChange}\\
				\noindent{\footnotesize{Symbols: \Acrfull{musp}, \acrfull{dmua}, optical property recovered with the optode displaced from the nominal position ($(\times\rho)$ in superscript)}}
			}
		\end{figure}

		The figures in \autoref{sec:RESsc} and \autoref{sec:RESds} reveal information about which optode displacement direction causes the largest error and which direction causes no change in the error.
		If an optode moves along its iso-line (i.e., line of constant error from the nominal optode position) in these figures, the recovered value will not change.
		Therefore, we wish to design a probe which prevents optode movement in directions which introduce errors but allows movement in directions that will not change the recovered value.
		\autoref{fig:linesNoChange} shows these lines of zero change for the LINR and TRAP arrangements.
		These lines show positions where the optode can be placed that will not change the recovered value of a particular optical property.
		The lines for different optical properties intersect at the nominal optode position since this is the only position which will cause no error change for any of the optical properties.
		We can use \autoref{fig:linesNoChange} to determine which directions it is allowable for optodes to move (i.e., along the lines of zero change) in a real-life probe design.
		\par
		
		\glsreset{mua}
		
\begin{figure}[H]
			\includegraphics{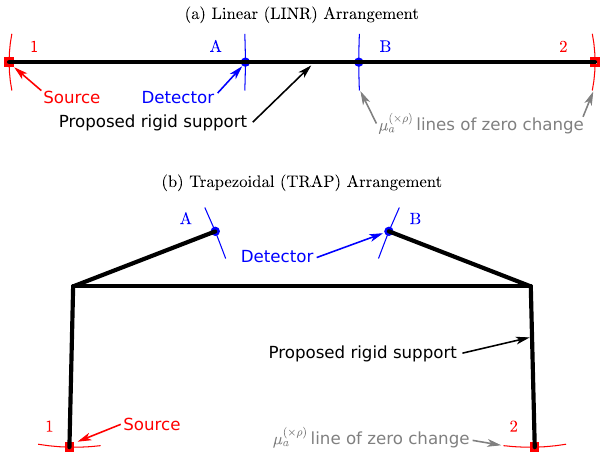}
			\caption{Proposed support structure which was designed by allowing moving along the lines of zero change (\autoref{fig:linesNoChange}) but restricting movement perpendicular to the line. Lines in this figure considered zero change in \gls{mua}.
				(a) The LINeaR (LINR) arrangement.
				(b) The TRAPezoidal (TRAP) arrangement.
				\label{fig:supports}\\
				\noindent{\footnotesize{Symbols: Optical property recovered with the optode displaced from the nominal position ($(\times\rho)$ in superscript)}}
			}
		\end{figure}

		\autoref{fig:supports} shows how these lines of zero change can be used to design a \gls{DS} / \gls{SC} probe.
		Since movement along the line of zero change is allowable but movement perpendicular to that causes an error, one can design rigid supports which are perpendicular to the tangent of the line of zero change.
		We then extend these supports until they intersect to show the proposed skeleton of rigid supports.
		This proposed support structure should reduce errors induced by incorrect assumptions of the optode position, since they do not allow the optodes to move in the direction of maximum error.
		For the LINR arrangement (\autoref{fig:supports}(a)) these supports are trivial since they run along the linear line of the probe.
		However, for the TRAP arrangement (\autoref{fig:supports}(b)), this method of designing supports results in a non-obvious structure.
		This approach to design \gls{SC} / \gls{DS} probes will result in measurements more robust against optode position errors and probe deformation.
		\par
		
		\glsreset{mua}
		
		%%%%%%%%%%%%%%%%%%%%%%%%%%%%%%%%%%%%%%%%%%
		\section{Conclusions}
		Self-calibrating and dual-slope measurements present a number of significant practical advantages by not requiring calibration of source emission and detector sensitivity \cite{Hueber_Opt.Tomogr.Spectrosc.TissueIII99_NewOptical}, and by being relatively insensitive to movement artifacts \cite{Scholkmann_PM14_MeasuringTissue}. 
		These measurements rely on the fulfillment of specific geometrical requirements for the location of optodes \cite{Blaney_Rev.SI20_DesignSource, Fantini_JIOHS19_TransformationalChange, Hueber_Opt.Tomogr.Spectrosc.TissueIII99_NewOptical} and knowledge of the location of all optodes.
		This work investigated the impact of uncertainties in the location of the optodes, which is especially relevant in the case of \textit{in vivo} applications to biological tissues where the optical probe is required to conform to the shape of the investigated tissue. 
		The significance of this work is twofold. 
		First, it allows one to estimate the impact of uncertainties in the optode positions on absolute or relative optical measurements from continuous-wave or frequency-domain data in dual-slope configurations. 
		Second, it guides the design of optical probes to minimize probe deformation along directions that result in optode displacements with greatest impact on the optical measurements.
		\par
		
		%%%%%%%%%%%%%%%%%%%%%%%%%%%%%%%%%%%%%%%%%%
		\vspace{6pt} 
		
		%%%%%%%%%%%%%%%%%%%%%%%%%%%%%%%%%%%%%%%%%%
		\authorcontributions{Conceptualization, G.B., A.S. and S.F.; methodology, G.B. and A.S.; software, G.B. and A.S.; validation, G.B. and A.S.; formal analysis, G.B. and A.S.; investigation, G.B., A.S., and T.D.; resources, S.F.; data curation, G.B., A.S., and T.D.; writing---original draft preparation, G.B., A.S., T.D., and S.F.; writing---review and editing, G.B., A.S. and S.F.; visualization, G.B.; supervision, A.S. and S.F.; project administration, S.F.; funding acquisition, S.F. All authors have read and agreed to the published version of the manuscript.}
		
		\funding{This work is supported by \gls{NIH} awards R01-NS095334 and R01-EB029414. G.B. would also like to acknowledge support from \gls{NIH} award K12-GM133314. The content is solely the authors' responsibility and does not necessarily represent the official views of the awarding institutions.}
		
		\dataavailability{Supporting code for this manuscript can be found at the following link:\\ \url{https://github.com/DOIT-Lab/DOIT-Public/tree/master/SelfCalibratingPositionErrors}} 
		
		%\acknowledgments{In this section you can acknowledge any support given which is not covered by the author contribution or funding sections. This may include administrative and technical support, or donations in kind (e.g., materials used for experiments).}
		
		\conflictsofinterest{The authors declare no conflicts of interest.} 
		
		%%%%%%%%%%%%%%%%%%%%%%%%%%%%%%%%%%%%%%%%%%
		%%\isPreprints{}{% This command is only used for ``preprints''.
			%\begin{adjustwidth}{-\extralength}{0cm}
			%%} % If the paper is ``preprints'', please uncomment this parenthesis.
		%%\printendnotes[custom] % Un-comment to print a list of endnotes
		
		\reftitle{References}
		
		
		\bibliography{GBlib250428}
		
%		\PublishersNote{}
		%%\isPreprints{}{% This command is only used for ``preprints''.
			%\end{adjustwidth}
			%%} % If the paper is ``preprints'', please uncomment this parenthesis.
	\end{document}